\documentclass{article}

\usepackage{arxiv}

\usepackage[utf8]{inputenc} 
\usepackage[T1]{fontenc}    
\usepackage[hidelinks]{hyperref}       
\usepackage{url}            
\usepackage{booktabs}       
\usepackage{amsfonts}       
\usepackage{nicefrac}       
\usepackage{microtype}      
\usepackage{graphicx}
\usepackage[numbers]{natbib}
\usepackage{doi}
\usepackage[table]{xcolor}

\usepackage{flafter}
\usepackage{layouts}
\usepackage{siunitx}
\usepackage[acronym, nogroupskip]{glossaries}
\usepackage{amsmath}
\usepackage{amssymb}
\usepackage{bbm}
\usepackage{subcaption}
\usepackage{mathtools,algorithm}
\usepackage[noend]{algpseudocode}
\usepackage{multicol}
\usepackage{longtable}
\usepackage{ltablex,booktabs}
\usepackage{cleveref}       
\usepackage{authblk}
\usepackage{changepage}

\algnewcommand{\IfThenElse}[3]{
	\State \algorithmicif\ #1\ \algorithmicthen\ #2\ \algorithmicelse\ #3}

\DeclareMathOperator*{\argmax}{arg\,max}
\DeclareMathOperator*{\length}{length}

\makeglossaries
\setacronymstyle{long-short}
\renewcommand{\glossarysection}[2][]{}

\newacronym{LWIR}{LWIR}{Long-wavelength infrared}
\newacronym{PCA}{PCA}{Principal component analysis}
\newacronym{KLT}{KLT}{Kanade-Lucas-Tomasi}
\newacronym{SVD}{SVD}{Singular value decomposition}
\newacronym{RoI}{RoI}{Region of Interest}
\newacronym{FFT}{FFT}{Fast Fourier Transform}
\newacronym{CWT}{CWT}{Continous Wavelet Transform}
\newacronym{RR}{RR}{Respiratory Rate}
\newacronym{RMSE}{RMSE}{Root Mean Square Error}
\newacronym{WFDB}{WFDB}{Waveform Database}
\newacronym{RE}{RE}{Respiratory Effort}
\newacronym{IRT}{IRT}{Infrared Thermography}
\newacronym{NIR}{NIR}{Near Infrared}
\newacronym{SVM}{SVM}{Support Vector Machine}

\title{Breathing Pattern Monitoring using Remote Sensors}

\date{19 September 2022}

\author{\href{https://orcid.org/0000-0002-9399-706X}{Janosch Kunczik \includegraphics[scale=0.06]{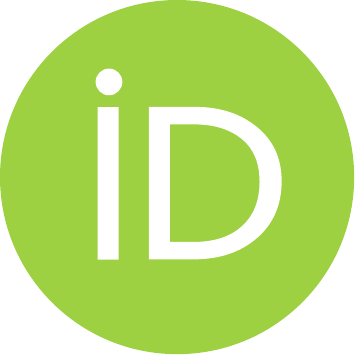} $^{1}$}}
\author{Kerstin Hubbermann $^{1}$}
\author{Lucas Mösch $^{1}$}
\author{\href{https://orcid.org/0000-0001-6817-2472}{Andreas Follmann \includegraphics[scale=0.06]{orcid.pdf} $^{1}$}} \author{\href{https://orcid.org/0000-0002-7118-7728}{Michael Czaplik \includegraphics[scale=0.06]{orcid.pdf} $^{1}$}} \author{\href{https://orcid.org/0000-0003-1788-4562}{Carina Barbosa Pereira \includegraphics[scale=0.06]{orcid.pdf} $^{1}$}}
\affil{$^{1}$ Department of Anesthesiology, RWTH Aachen University, Faculty of Medicine, Aachen, Germany}

\renewcommand{\headeright}{preprint}
\renewcommand{\undertitle}{preprint}
\renewcommand{\shorttitle}{Breathing Pattern Monitoring using Remote Sensors}

\hypersetup{
pdftitle={Breathing Pattern Monitoring using Remote Sensors},
pdfsubject={eess.SP, eess.IV},
pdfauthor={Janosch Kunczik, Kerstin Hubbermann, Lucas Mösch, Andreas Follmann, Michael Czaplik, Carina Barbosa Pereira},
pdfkeywords={contactless, classification, infrared thermography, RGB video, dataset, signal, extraction},
}

\begin{document}
\maketitle

\begin{abstract}
Breathing is one of the most important body functions because it provides it with oxygen, which is vital for energy production. In addition, the removal of carbon dioxide actively regulates the acid-base level, which is essential for the physiological function of the body. Due to its close connection with many other body functions, respiration can also be used as an indicator for a wide spectrum of medical conditions, which at first glance have little to do with breathing. Neurological, cardiological, inflammatory, metabolic, and even psychological conditions symptomatically show up in breathing patterns. Hence, being able to classify them automatically and unobtrusively, can allow cost-effective monitoring systems to continuously assess the health of a patient. In this work, multiple respiratory signal-extraction algorithms for thermal and RGB cameras are presented and compared. A novel algorithm for the extraction of multiple respiratory features is presented and evaluated. Using a one vs. one multiclass support vector machine, these features were used to classify a wide range of respiratory patterns with an accuracy of up to \SI{95.79}{\percent}.
\end{abstract}

	
\section{Introduction}
Respiration provides the most important information to asses a patient for potentially life-threatening conditions. Various algorithms in emergency medicine first call for verification and assurance of adequate breathing, before other diagnostic or therapeutic steps are to be taken \cite{soar_2019_2019,the_atls_subcommittee_american_college_of_surgeons_committee_on_trauma_and_the_international_atls_working_group_advanced_2013,thim_initial_2012}. Even the decision whether to resuscitate an unconscious patient is made only on the basis of whether spontaneous breathing is present. Respiration is responsible for supplying the body with oxygen, which is essential for energy production and thus the maintenance of body functions. Furthermore, the respiratory activity regulates the $\text{CO}_2$ levels by removing it during expiration, which directly influence the body's pH-levels. Hence, disturbances in these processes are directly life threatening, as all other body functions rely on them. As result, many critical problems can be firstly observed in the respiratory activity, before they manifest in other vital signs, such as heart rate, peripheral oxygen saturation or blood pressure. This is the reason why breathing plays such a vital role in emergency medicine.

Assessing respiration can not only give insight about critically important gas exchange, but can also be used as an indicator for a wide range of medical conditions. Traumatic injuries, cardiac, neurological, inflammatory, metabolic and psychological conditions often lead to distinct changes in the breathing pattern. Therefore, an automatized monitoring of respiratory activity can give valuable information about the patient's health state and could be used to prevent a wide spectrum of fatalities by allowing an early detection of these conditions.

The current state of the art for breathing pattern assessment utilizes respiratory belts strapped around a subjects chest. While this approach is feasible for clinical research and sleep labs, it is not suitable for long-term patient monitoring due to cumbersome cables. Hence, contactless assessment of respiratory patterns would be desirable as it would allow monitoring a broad range of patients on ordinary wards in hospitals, in nursing homes and even at home for possible life threatening conditions.

\subsection{State of the Art and Scope of this Work}
Multiple aproaches have been published, which allow a contactless extraction of the \gls{RR} \cite{al-khalidi_respiration_2011}. Most of the approaches found in the literature utilize RF-based methods, such as radar, WiFi, or software defined radios. Camera based approaches using RGB, \gls{IRT}, and \gls{NIR} are also very common. However, apart from some pure frequency considerations, the classification of different breathing patterns has not been shown, yet. Rehman et al. \cite{rehman_improving_2021} stated that a respiratory pattern classification based on software defined radio measurements is possible and presented classifiers with good accuracy. However, as the description of the study protocol was kept very brief and the recorded dataset was not published, it can not be confirmed that the good classifications results are not mostly based on a sufficient separation in frequency between the different patterns and overfitting. Romano et al.\cite{romano_non-contact_2021} and Barbosa Pereira et. al. \cite{barbosa_pereira_estimation_2017} both performed a study in which multiple subjects were asked to breath following a on-screen animation with different respiratory patterns. The experiments were recorded with RGB and \gls{IRT} cameras, respectively. The data was then used to validate and compare multiple respiratory signal extraction algorithms against a gold standard by means of \gls{RR} extraction quality. However, a classification of the different breathing patterns was not tried. Additionally,while many works in the past have extracted respiratory signals from \gls{IRT}-, RGB- videos, or a combination of both, to the authors' knowledge, there is no work that directly compares the signal qualities obtained from both modalities. 

Therefore, this work presents the results of a subject study, which was conducted similarly to the aforementioned studies, but while capturing the experiments with both, \gls{IRT}, RGB cameras and respiratory belts. The recordings were augmented to cover the whole range of expected respiratory patterns, in order to prevent overfitting.  Subsequently, different signal extraction algorithms of all modalities were compared against each other to determine the best modality and extraction algorithm for contactless respiratory pattern classification. Furthermore, a robust extraction algorithm for multiple respiratory features is presented and evaluated. Lastly, a classifier is proposed, which is capable of distinguishing between all respiratory patterns with an accuracy of up to \SI{95.79}{\percent}.

\subsection{Respiratory Patterns}
In particular, medical textbooks contain descriptions of all breathing patterns (see e.g. \cite{whited_abnormal_nodate, brandes_physiologie_2019, ricard_pulmonary_2014, olivencia_pena_spontaneous_2020}). However they often contradict each other, so that this section shall be used to define and motivate the breathing patterns discussed in this publication.

\begin{figure}[htpb]
	\centering
	\includegraphics[]{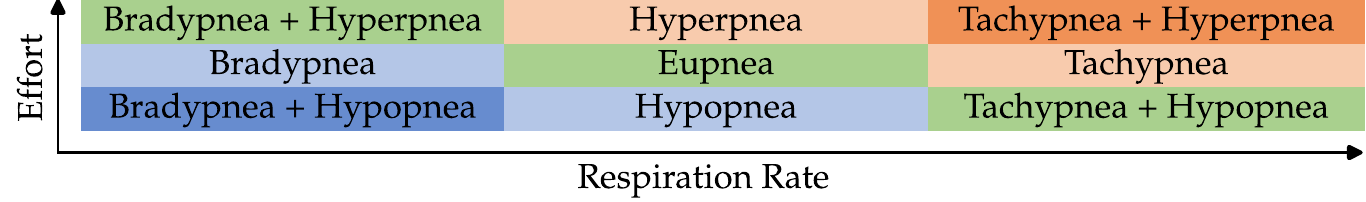}
	\caption{Qualitative respiration rate- and depth-based pattern differentiation: Hyperventilation and hypoventilation are marked by saturations of orange and blue, respectively. The more saturated the color, the more pronounced the condition. } 
	\label{fig:NaivePatterns}
\end{figure}

Normal respiration is called eupnea and is defined by nearly sinusoidal thorax movements with \gls{RR} between 12 and 18 \si{Breaths\per\minute} \cite{noauthor_atemfrequenz_nodate}. Tachypnea, which is characterized through higher \gls{RR} than for eupnea, can be caused by either an increased oxygen demand (e.g. during exercise or fever), a reduced gas exchange capability (e.g. through asthma, pneumonia, pulomonary embolisms, etc.), or an oversupply of $\text{CO}_2$ through e.g. metabolic acidodsis. The reasons for tachypnea can also cause a similar phenomenon, in which instead with increased frequency the body copes with an increased \gls{RE} (deeper breaths), which is then called hyperpnea. Because the three patterns are closely linked to each other,  Hyperpnea is often synonymously referred to as Kussmaul breathing or hyperventilation, which is not entirely correct. Following the original publication from Kussmaul \cite{kussmaul_zur_1874}, his discovered pattern is defined by deep, laboured breaths with increased frequency, which can be seen in patients with severe metabolic acidosis. Hence, Kussmaul breathing is a combination of hyperpnea and tachypnea. Hyperventillation, often a sign for psychological conditions, like anxiety is defined through a pathologically high gas exchange, which can be caused by tachypnea, hyperpnea or a combination of both. Figure \ref{fig:NaivePatterns} shows this interaction between different breathing patterns. Hypoventillation on the other side, is often caused by bradypnea, or hypopnea, which are defined by lower \gls{RR}s or \gls{RE}s, respectively. However, in contrast to hyperventilation, it can be also caused a reduced gas exchange an manifest itself in tachy- or hyperpnea. Hence, there are exceptions to the color scheme in Figure \ref{fig:NaivePatterns}. 

Breathing is controlled by the respiratory center, which is located in the brain stem as part of the autonomic nervous system. Because the respiratory center is located near the narrow cranio-cervical junction, where the spinal cord leaves the cranial cavity, it can be disturbed by increased intracranial pressure, traumatic brain injuries and through Meningism (inflammation of the protective membranes of the spinal cord). Intoxications can also influence its functions. All the aforementioned conditions can reduce its activity and thus lead to bradypnea, and hypopnea which can even develop into respiratory arrest (apnea). Strong impairments of the respiratory center manifest themselves as a pathological pattern called Biot's breathing, which is characterized by periods of sufficent breathing, which are randomly interrupted  with phases of apnea. An oxygen deficiency of the brain stem, for example due to a stroke or heart attack can yield a breathing pattern called Cheyne-Stokes which is like Biot's breathing defined by intermitted phases of apnea, but with increasing and decreasing \gls{RE}s. A visual representation of all discussed respiratory patterns can be found in Figure \ref{fig:breathPatterns}. Since this work aims to classify respiratory patterns by video-based approaches, which can only capture changes in \gls{RR} and \gls{RE}, ventillation based patterns, like hyper- and hypoventilation weren't further discussed.

\section{Materials and Methods}
\subsection{Data Collection}
Training and validation data was collected by means of a voluntary subject study with 10 male and female subjects aged 24 to 45 years. During a 15 minutes long recording, all subjects were asked to sit comfortably in the provided chair and to breathe according to the predefined breath-patterns. In advance, each subject was informed about the exact purpose and procedure of the study. Furthermore, it was expressly pointed out that the test protocol only needs to be followed at one's own discretion. All subjects were informed that they could interrupt the experiment at any time, if e.g. breathing patterns were causing discomfort of any kind. The study was conducted after notification of the Ethics Committee of the University Hospital RWTH Aachen University.

\begin{figure}[htpb]
	\centering
	\includegraphics[width=\textwidth]{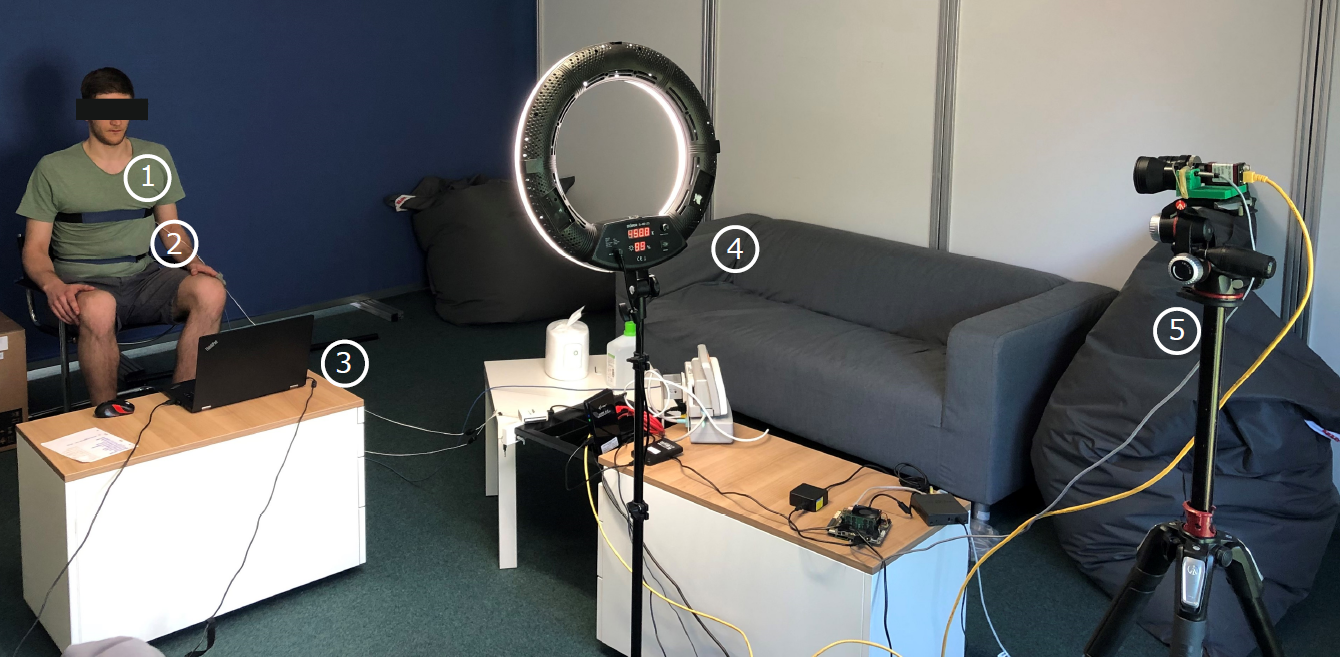}
	\caption{Experimental setup: A voluntary subject is sitting in a chair, wearing two \gls{RE} sensors around thorax (1) and abdomen (2). A computer (3) displays predefined breathing patterns and stores the video frames of an RGB and LWIR camera (5). Illumination is ensured using a ring light (4).}
	\label{fig:setup}
\end{figure}

Data was collected as shown in Figure \ref{fig:setup} using three different sensor modalities. Two respiratory belt sensors, placed around thorax (1) and abdomen (2) of the subject served as gold standard reference signal sources. A RGB- and a \gls{IRT} camera (5) were used to record the upper body of the sitting subject. To ensure proper illumination, a ring light (4) was placed between the subject and the cameras. Animations of the pre-defined breathing patterns were displayed on a computer (3) in front subject. MATLAB \cite{moler_matlab_2022} was the software used for data collection and all evaluation. Detailed information about the used devices can be found in \ref{tab: caption_devices}.

\begin{figure}[htpb]
	\centering
	\includegraphics[]{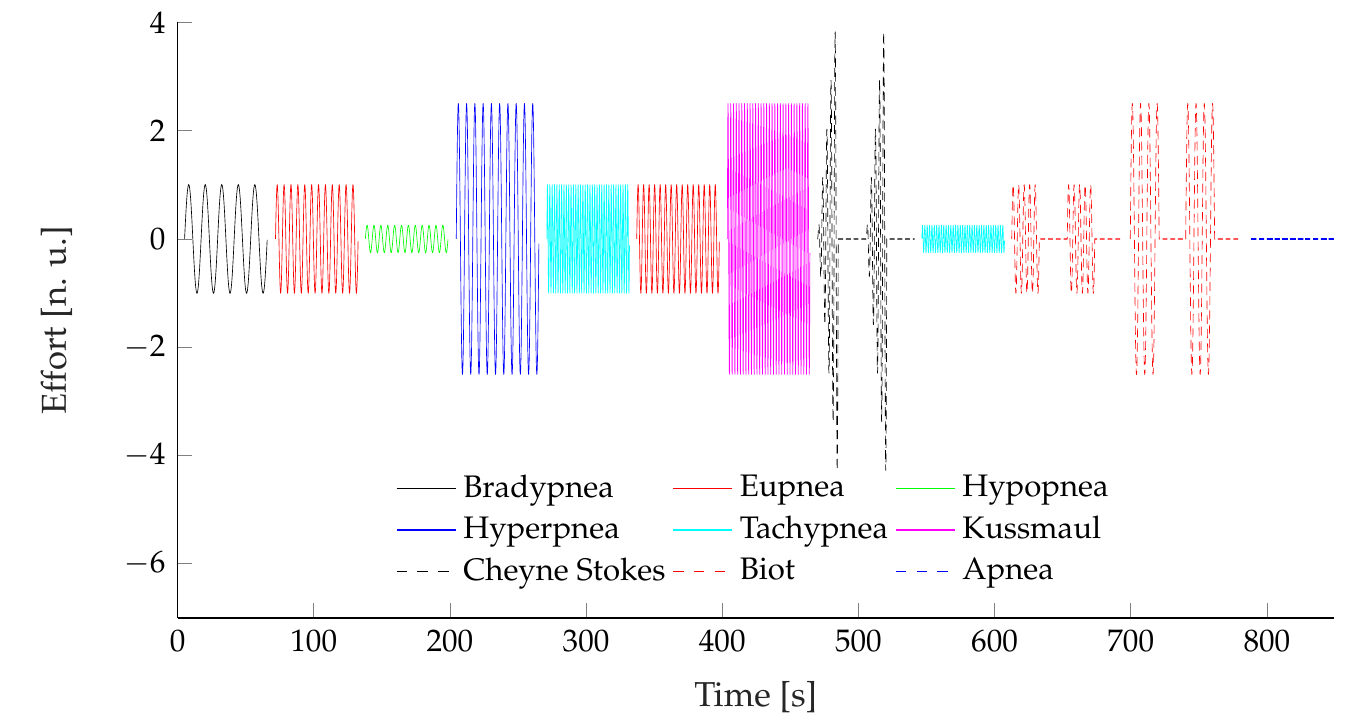}
	\caption{Visual representation of the breathing-pattern protocol.} 
	\label{fig:breathPatterns}
\end{figure}

The sequence of predefined breathing patterns is shown in Figure \ref{fig:breathPatterns}. It is henceforth referred to as \textbf{reference} signal, as it was used to compare all other modalities against it. A detailed description of the used patterns can be found in table \ref{tab:breathPatterns}. The protocol was designed and tested to alternate the different patterns such that, on average a normal gas exchange occurs. After each pattern, a five second pause was added, in which the subjects could breathe freely. 

\subsection{Reference-signal Extraction}
The readings from chest and abdomen belt sensors were over-sampled with \SI{5}{\kilo\hertz}, digitally low-pass filtered using a moving mean filter and then down-sampled to $f_s=\SI{10}{\hertz}$. The extracted signals, from here on referred to as \textbf{chest belt} and \textbf{abdomen belt}, were used as gold standard references to determine the performance of their contactlessly extracted counterparts.

\subsection{Raw RGB Signal Extraction} \label{sec:RGBExtraction}
\begin{figure}[htpb]
	\centering
	\includegraphics[]{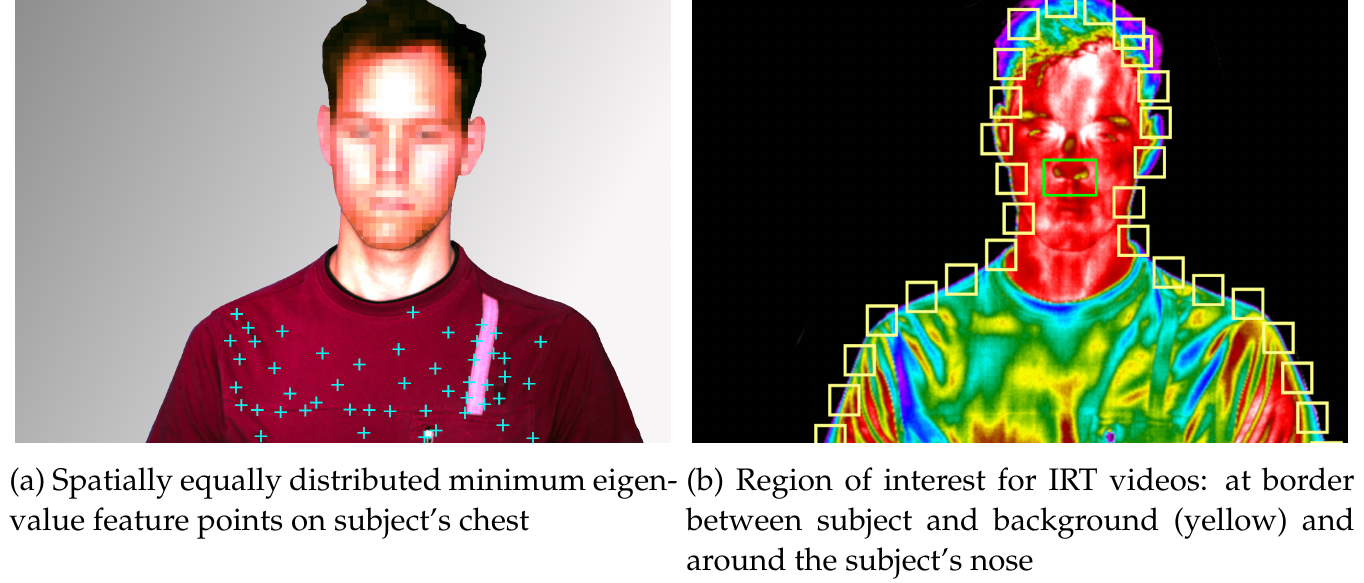}
	\caption{Points and regions of interest for respiratory signal extraction for color and thermal videos.} 
	\label{fig:RoIs}
\end{figure}

Respiratory signal extraction from RGB footage was achieved by tracking the motion subjects chest. For this purpose, 50 feature points were selected from this region, using the minimum eigenvalue algorithm \cite{jianbo_shi_good_1994}. These points were used to set-up and run a \gls{KLT} tracker \cite{lucas_iterative_1981, tomasi_detection_1991} (see Figure~\ref{fig:RoIs}a), which tracked their positions in cartesian coordinates over the full length of the video. To transform the $n$ individual, two-dimensional trajectories $\mathbf{T}_i \in \mathbb{R} ^{m\space x2}, i \in [1, n]$ with length $m$ into one-dimensional signals, a \gls{SVD} was utilized: 
\begin{equation}
	\mathbf{T}_i = \mathbf{U}_i \mathbf{\Sigma}_i \mathbf{V^*}_i.
\end{equation}
The \gls{SVD} is a generalized form of the eigenvalue decomposition for non-square matrices and uses two change-of-basis matrices $\mathbf{U}_i \in \mathbb{R} ^{2\space x\space 2}$, $\mathbf{V^*}_i \in \mathbb{R} ^{m\space x\space m}$ to rotate the signal's vector space, such that the semi-major and semi-minor axes of an ellipse, fitted to the signal distribution align with the coordinates systems base vectors. In the existing case, $\mathbf{U}_i$ and $\mathbf{V^*}_i$ represent base change matrices for the spatial and temporal domains, repsectively. This allows for an reduction of the trajectories spatial dimensionality to only their main direction of motion, by mutliplying them with the first column vector of $\mathbf{U}_i$, called $\vec{u}_{i,1}$:
\begin{equation}
	\vec{o}_i = \mathbf{T}\cdot \vec{u}_{i,1}.
\end{equation}
The results were collected in the observation matrix $\mathbf{O} \in \mathbb{R} ^{m\space x\space n}$
\begin{equation}
	\mathbf{O} = 	\begin{bmatrix}
		\vec{o}_1 & ... & \vec{o}_i & ... & \vec{o}_n
	\end{bmatrix},
\end{equation}
which is composed by $n$ trajectories $\vec{o}_i \in  \mathbb{R} ^{m}, \space i \in [1,n]$ with length m. These signals are referred to in the following as \textbf{chest RGB}. 

\subsection{Raw Thermal Signal Extraction}
The respiratory signal extraction from thermal videos was performed using three independent approaches: 
\begin{itemize}
	\item observation of temperature changes below the nostrils (referred to as \textbf{nose \gls{IRT}}),
	\item tracking the subject's upper body movement in contrast to the background (referred to as \textbf{border RoI}) and
	\item tracking of feature point on the subjects chest, analog to the RGB-based approach in section \ref{sec:RGBExtraction} (referred to as \textbf{chest \gls{IRT}}).
\end{itemize}
The first approach was investigated in multiple publications \cite{pereira_remote_2015, pereira_robust_2015, mutlu_ir_2018} and is based on the fact that inspiration cools and expiration warms the nostrils. Hence, their temperature change can be used as respiratory signal. To extract the nostril temperature, a \gls{RoI} around the subject's nose was selected (see Figure \ref{fig:RoIs}b) and tracked using minimum eigenvalue \cite{jianbo_shi_good_1994} feature points in the subject's face and a \gls{KLT} tracker \cite{lucas_iterative_1981, tomasi_detection_1991} (see Figure \ref{fig:RoIs}). The \gls{RoI}'s mean temperature was extracted in every frame and was directly used as respiratory signal $s[k]$.

For the second approach, the high natural contrast between the subject and the background was used. The first recorded thermal frame was automatically segmented into forground and background, using Otsu's method \cite{otsu_threshold_1979}. Subsequently, all regions in the foreground mask were ordered by their area. The region with the largest area was treated as a mask for the test subject. To extract respiratory signals, a number of $n = 30$ \gls{RoI}s with a size of 25 x 25 pixels were initialized along the border of the test subject and the background (see Figure \ref{fig:RoIs}b). Their mean value was computed for every frame of the video and again saved in an observation matrix $\mathbf{O}$. To make the signal extraction robust against larger movements of the test subject, the two-norm of the difference of all consecutive frames was computed:
\begin{equation}
	e_{m,k} = ||\mathbf{O_k}-\mathbf{O_{k-1}}||^2.
\end{equation}
If the motion error $e_{m,k}$ grew larger than 5, the frame was marked as unusable and discarded. Two seconds after the latest, marked frame, all \gls{RoI}s were re-initialized and the signal extraction was continued.

\subsection{Robust Respiratory Signal Extraction}
Not all signals in the observation matrices $\mathbf{O}$ are mainly influenced by the subject's respiratory activity, but may also contain motion artifacts or only consist of noise. These signals can not be used for a robust respiratory signal extraction an thus should be discarded. To decide which signals are suitable, the correlation matrix $\mathbf{P}$ of $\mathbf{O}$ was computed. It was assumed that the respiratory activity yields a high correlation between all signals in which it can be observed. Thus, the mean correlation between a signal and all other observed signals was calculated:

\begin{equation}
	\rho_{\mu,i} = \sum_{j=1}^{n} \frac{\rho_{j,i}}{n}, i \in [1, n].
\end{equation}

Only the five signals with the biggest mean correlation coefficient were kept in a reduced observation matrix $\mathbf{O}_{max}$, which was then used as input for a moving window \gls{PCA} with a window length $w_l$ of \SI{30}{\second}:
\begin{equation}
	\mathbf{C}_k = \begin{bmatrix}
		\vec{c}_1, & ..., & \vec{c}_t 
	\end{bmatrix} = PCA(\mathbf{W}_k \cdot \mathbf{O}_{max}).
\end{equation}
The window matrix $\mathbf{W}_k = \mathbbm{1}_{\{1, ..., k-1, k + w, .., m\}} \in  \mathbb{R} ^{w_l, n}$ is a submatrix of a m-dimensional identity matrix, which only selects the observation points $[k, k+w_l-1]$ and thus only provides a portion of the observation matrix to the \gls{PCA} algorithm. 

A \gls{PCA} is an orthogonal, linear, transformation, which transforms the data into a new coordinate systems, where the first axis corresponds to the direction of the greatest signal variance. Since the variance of a signal is independent of its sign, the transformation-coefficients $\mathbf{C}$ may change their sign for every move of the window, which causes discontinuities in the result. To resolve this problem, the signs of the current coefficient matrix $\mathbf{C}_{k}$ were adjusted, to be identical to its predecessor, using the direction matrix $\mathbf{D}_k$:
\begin{equation}
	\mathbf{D}_k = sgn(\mathbf{C}_{k}\circ\mathbbm{1})\cdot sgn(\mathbf{C}_{k-1}\circ\mathbbm{1}),
\end{equation}
where $\circ$ is the Hadamard product. Finally, the robust respiratory signal $s[k]$ at time k could be computed by multiplying the observation vector $\vec{o}_k$ with the direction matrix $\mathbf{D}[k]$ and the first column vector $\vec{c}_{1,k}$ of the coefficent matrix $\mathbf{C}[k]$, which represents the signal's principal component:
\begin{equation}
	s[k] = \vec{o}_k\cdot\mathbf{D}_k\cdot\vec{c}_{1,k}.
\end{equation}

\subsection{Dataset Creation}
After the respiratory signal extraction from all aforementioned modalities and approaches, the sequences were manually labeled by comparing them with the reference signal. Because for different modalities and different subjects, the signal amplitudes varied noticeably, all signals were normalized. For this purpose, the signal amplitudes in phases with apnea were calculated by finding the maximum amplitude in their \gls{FFT} spectrum. These amplitudes were defined to have the value 1 normailzed unit (n.u.):
\begin{equation}
	s_{norm}[k] = \frac{1}{max(|\mathfrak{F}(\vec{s}_{eup})|)}\cdot s[k].
\end{equation}
It was noted that all signals exhibited a slight systematic error in their sampling rate. Therefore, all signals of a modality were resampled to match the desired sampling rate and were then aligned with each other by finding the most likely delay $d_{i,j}$ between every signal pair, using their cross correlation. Using a least squares optimization problem of the form
\begin{equation}
	\begin{bmatrix}
		d_{1,2}\\
		\vdots \\
		d_{i,j}\\
		\vdots \\
		d_{n-1,n}\\
	\end{bmatrix} = 
	\begin{bmatrix}
		1 & -1 &  & \cdots &  &  & 0 \\
		&  &  & \vdots  &   &  &  \\
		0 & \cdots & 1 & \cdots & -1 & \cdots & 0 \\
		&  &  & \vdots  &   &  &  \\
		0 &  &  & \cdots &  & -1 & 1 \\
	\end{bmatrix}
	\begin{bmatrix}
		d_{1}\\
		\vdots \\
		d_{i}\\
		\vdots \\
		d_{n}\\
	\end{bmatrix},
\end{equation}
these distances were converted into the specific signal offests $d_{i}$. Some signals exhibited a larger error in their sampling rate and were discarded in the first resampling and alignment step. To match those signals with the rest, a grid search over all possible resampling factors was performed. The factors that yielded the highest cross correlation value and their corresponding lag were chosen to resample and delay these signals accordingly. The dataset was then stored in the MIT \gls{WFDB} format.

\begin{figure}[htpb]
	\centering
	\includegraphics[]{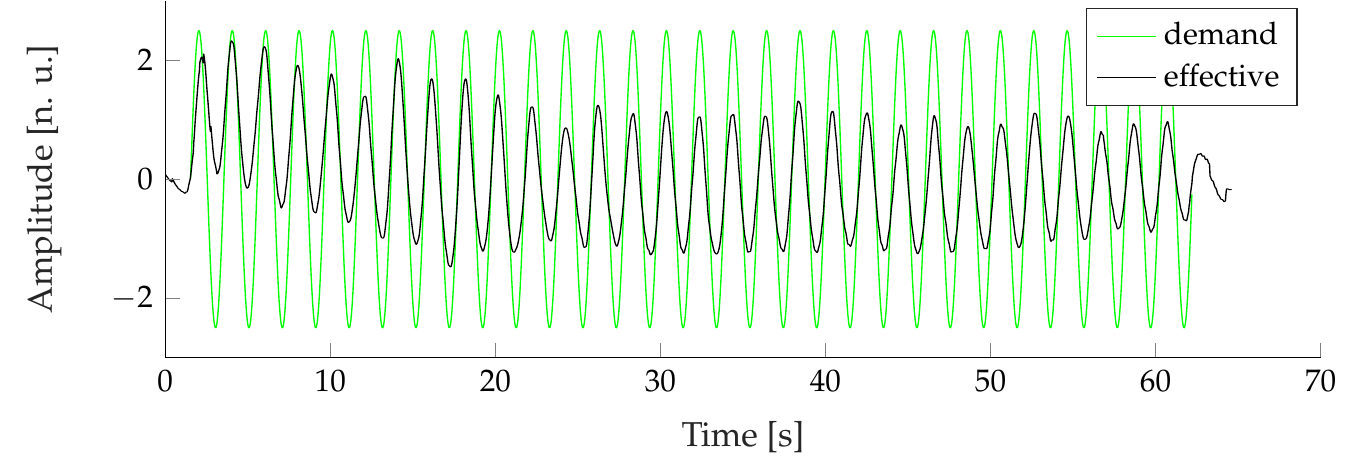}
	\caption{Kussmaul pattern excluded from dataset, because of big difference between demanded respiratory activity and the mean effective respiratory response of all subjects.} 
	\label{fig:KussmaulDismissal}
\end{figure}

To augment the dataset, the original data was altered by combining arbitrary signals $s_i[k]$ and $s_j[k]$ of different subjects, with each other using random percentages $p\,\in\,[0, 1]$:
\begin{equation}
	s_{\text{gen},l}[k] = p\cdot s_i[k] + (1 - p)\cdot s_j[k].
\end{equation}
Afterwards, the signals $s_{\text{gen},l}[k]$ were split into the individual, recorded breathing patterns. For each breathing pattern, the signal was either stretched, or compressed to randomly vary the \gls{RR} within its specified target range (see \ref{tab:breathPatterns}). Through this approach, the original dataset size was increased to 1000 samples for each pattern, modality and extraction approach, to allow the utilization of arbitrary  machine learning classifiers. Furthermore, the \gls{RR} of the recorded signals was randomly distributed to prevent classifiers from simply distinguishing several patterns through their distinct frequency. Because the required \gls{RE} of the recorded Kussmaul breathing pattern could not be replicated by any subject (see Figure \ref{fig:KussmaulDismissal}), those sections were discarded and replaced with hyperpnea patterns that were shifted into the higher Kussmaul \gls{RR} range.

\subsection{Feature Extraction}
In a pre-processing step, the respiratory signals $s[k]$ were smoothed by filtering with a Savitzky-Golay filter \cite{savitzky_smoothing_1964} with the parameters shown in table \ref{tab:SgolayParams}. Using MATLABs  finpeaks algorithm, all peak locations $T_p[k]$, their amplitudes $a[k]_{p,\textrm{nu}}=0.5\cdot p[k]$ (where $p[k]$ is the peak prominence) and half-widths  $w[k]_{p,\textrm{nu}}$, which fulfilled the criteria given in table \ref{tab:FindPeaksParamas} were extracted from the smoothed signal. 

The \gls{RR} was computed through the inverse of the continued differentiation between two consecutive peak locations $T_{p,k}$ and $T_{p,k-1}$:
\begin{equation}
	RR_{p,\textrm{nu}}[k] = \frac{60}{T_{p,k} - T_{p,k-1}}.
\end{equation}
Because $RR_{p,\textrm{nu}}[k]$, $A_{p,\textrm{nu}}[k]$ and $w_{p,\textrm{nu}}[k]$ are non-uniformly sampled with the occurrence of the inspiratory peaks, they were re-sampled into $RR_{p}[k]$, $A_{p}[k]$ and $w_{p}[k]$ to match the length and sampling rate of $s[k]$, using a next neighbour interpolation, which holds the last detected value until the next value.

\begin{figure}[htpb]
	\centering
	\includegraphics[]{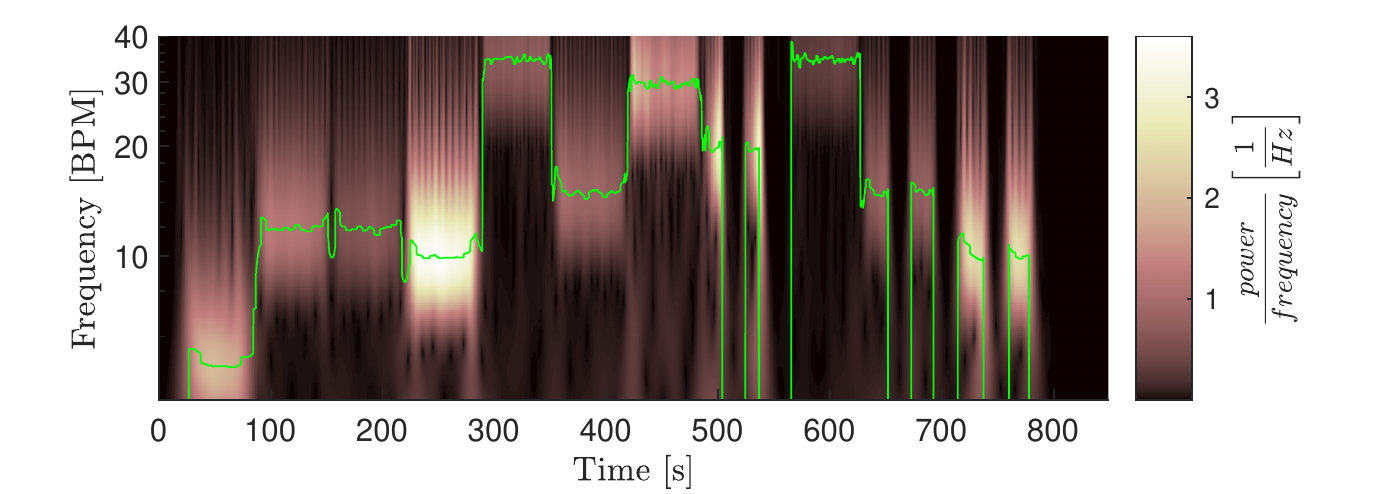}
	\caption{Continuous wavelet spectrum of the mean RGB signal, including the extracted Respiratory Rate.} 
	\label{fig:CWT}
\end{figure}

To increase the robustness of the feature extraction, the \gls{RR} and its amplitude were also extracted from the signal's spectrum using \gls{CWT} ridges. Ridge points represent the local maxima of a spectrogram and indicate a signal's instantaneous frequencies and amplitudes within the resolution of the used transform \cite{stephane_time_2009}. Hence, the \gls{RR} and the respiratory amplitude were computed from continuous wavelet spectrogram $S[f,k]$ of $s[k]$ by finding the frequency $f\, \forall\, k$, which maximizes the value of $S$:
\begin{equation}
	\begin{split}
		RR_S[k] & =  60\cdot\argmax_{f}(S[f,k]), \\
		A_S[k]  & = \max_{f}(S[f,k]).
	\end{split}
\end{equation}
The parameters used to compute $S[f,k]$ can be found in \ref{tab:CWTParamas}. A \gls{CWT} spectrogram with its ridge signal is shown in Figure~\ref{fig:CWT}.

\subsection{Artifact Correction} \label{sec:ArtifactCorrection}
Artifact correction was done on both feature tuples $F_p[k]=\left(RR_p[k], a_p[k], w_p[k]\right)$ and $F_s[k]=\left(RR_S[k], a_S[k]\right)$ individually. For the ease of notation, the following equations are thus derived from a generic feature set $F_{i}[k]\,i\in\,[p, S]$. 

Signals segments that were too discontinuous, showed noisy behavior below a certain respiratory amplitude, had a too low \gls{RR}, or contained to few valid values, were invalidated. To remove segments with too many discontinuities in \gls{RR}, all steps greater than \SI{5}{Breaths\per\minute} were marked. If two discontinuities occurred within a \SI{10}{\second} time frame, the associated section was discarded. 

Signals with amplitudes below 0.05 n.u. were marked as unreliable. In order to include the associated segments anyway, three conditions had to be met: the segment $F_{i,j}\,\subset\,F_{i}$ had to be longer than five times the period length of its median \gls{RR}, the median \gls{RR} had to be greater than \SI{5}{Breaths\per\minute} and the onto the median \gls{RR} normalized difference between the smallest and largest \gls{RR} had to be smaller than 0.45 n.u.:
\begin{equation}
	\begin{split}
		RR_{\textrm{med}} & = \widetilde{RR}_{i,j}[k], \\
		\length_{\textrm{in}\,[\si{\second}]}(F_{i,j}[k]) & \geq \frac{5\cdot 60}{RR_{\textrm{med}}}, \\
		RR_{\textrm{med}} & > 5\,\si{Breaths \per \minute}, \\
		\frac{\max({RR_{i,j}[k])} - \min({RR_{i,j}[k]})}{RR_{\textrm{med}}} & > 0.45.
	\end{split}
\end{equation}
Lastly, all segments, shorter than \SI{10}{\second}, were invalidated, as well. All invalid segments shorter than  \SI{10}{\second} were subsequently filled with a moving median interpolation of the last five valid values. Segments not satisfying this condition were set zo zero.

\subsection{Robust Feature Fusion}
\begin{figure}[htpb]
	\centering
	\includegraphics[]{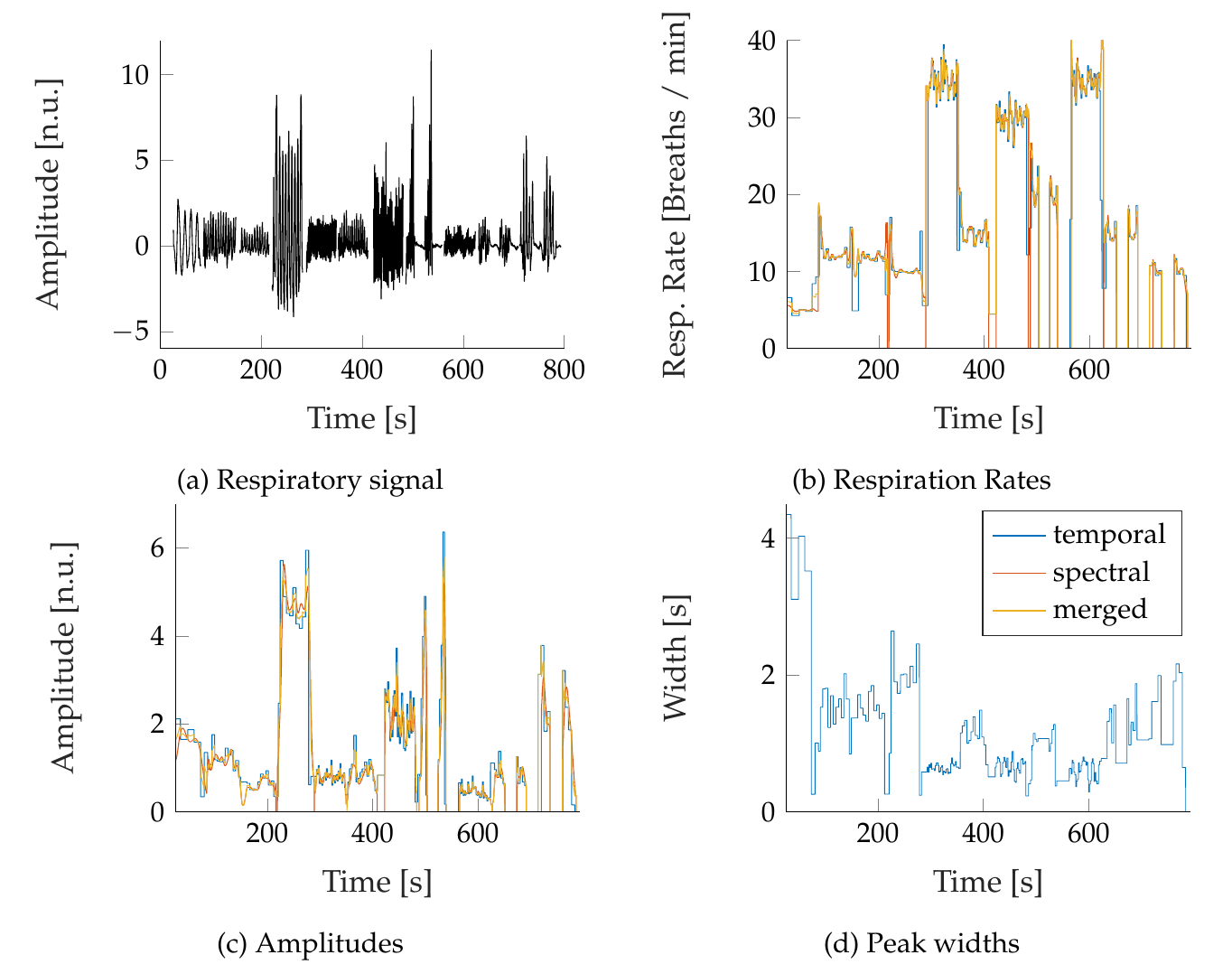}
	\caption{Time progression of the extracted features of an exemplary signal (RGB video of subject 6). All spectral, temporal and robustly merged features are displayed.} 
	\label{fig:TemporalFeaturesView}
\end{figure}

Figure \ref{fig:TemporalFeaturesView} shows the the peak- and \gls{CWT}-ridge-based feature sets next to its underlying respiratory signal (Figure \ref{fig:TemporalFeaturesView}a). Though both sets were artifact corrected, it can be seen that they still exhibit some artifacts, independently from each other (For example between 100 - 300 \si{\second} in Figure \ref{fig:TemporalFeaturesView}b). Generally, it has been observed that the temporally extracted feature set was more reliable during highly transient conditions, like during Biot's breathing and that the \gls{CWT}-ridge method performed better in low amplitude scenarios, like during tachy- and hypopnea. 

Hence, the results of both approaches were merged into a robust feature set. For this purpose, all signal jumps greater than \SI{5}{Breaths\per\minute} in the two respective \gls{RR} were marked and used as segment borders. For each segment the, signals were merged with the following rules: If the difference between the median \gls{RR} of both extraction approaches was smaller \SI{3}{Breaths\per\minute}, the mean of both signals was used. Else, the segment that had no signal jump in \gls{RR} for at least \SI{10}{\second} and whose mean \gls{RR} was higher compared to the other was used. If no valid section was present, the merged signal was set to an invalid value and later on filled using a moving median interpolation over the last 5 valid values and a maximum signal gap of \SI{5}{\second}. Values that were still invalid afterwards were set to zero. From the robust feature set, moving signal variances $RR_{\text{var}}$ and $A_{\text{var}}$ were computed with a time window of \SI{30}{\second} as a feature of change in the \gls{RR} and \gls{RE}.


\subsection{Classification}
For Breathing pattern classification a one vs. one multiclass \gls{SVM} classifier was utilized. The input to the classifier were chosen to be the features with  $RR_{\text{med}}$ and $A_{\text{med}}$ $RR_{\text{var, med}}$ and $A_{\text{var, med}}$ computed by calculating their median over complete breathing pattern sections. \gls{SVM}s  were first introduced by Wapnik et al. \cite{wapnik_theory_1974} and are binary classificators that optimize a linear border between two labeled classes, such that the distance of their datapoints to the border is maximized. To allow the classification of multiple classes, in the one vs. one approach proposed by Hastie et al. \cite{trevor_hastie_classification_1998}, a classifier for every combination of two classes is trained. The classifier was trained and validated using a 10-fold cross validation, in which the dataset was randomly split into 10 subsets, from which in every step 9 were used for training and one was used as validation data. 

\section{Results}
\subsection{Signal Extraction}
\begin{figure}[htpb]
	\centering
	\includegraphics[width=\textwidth]{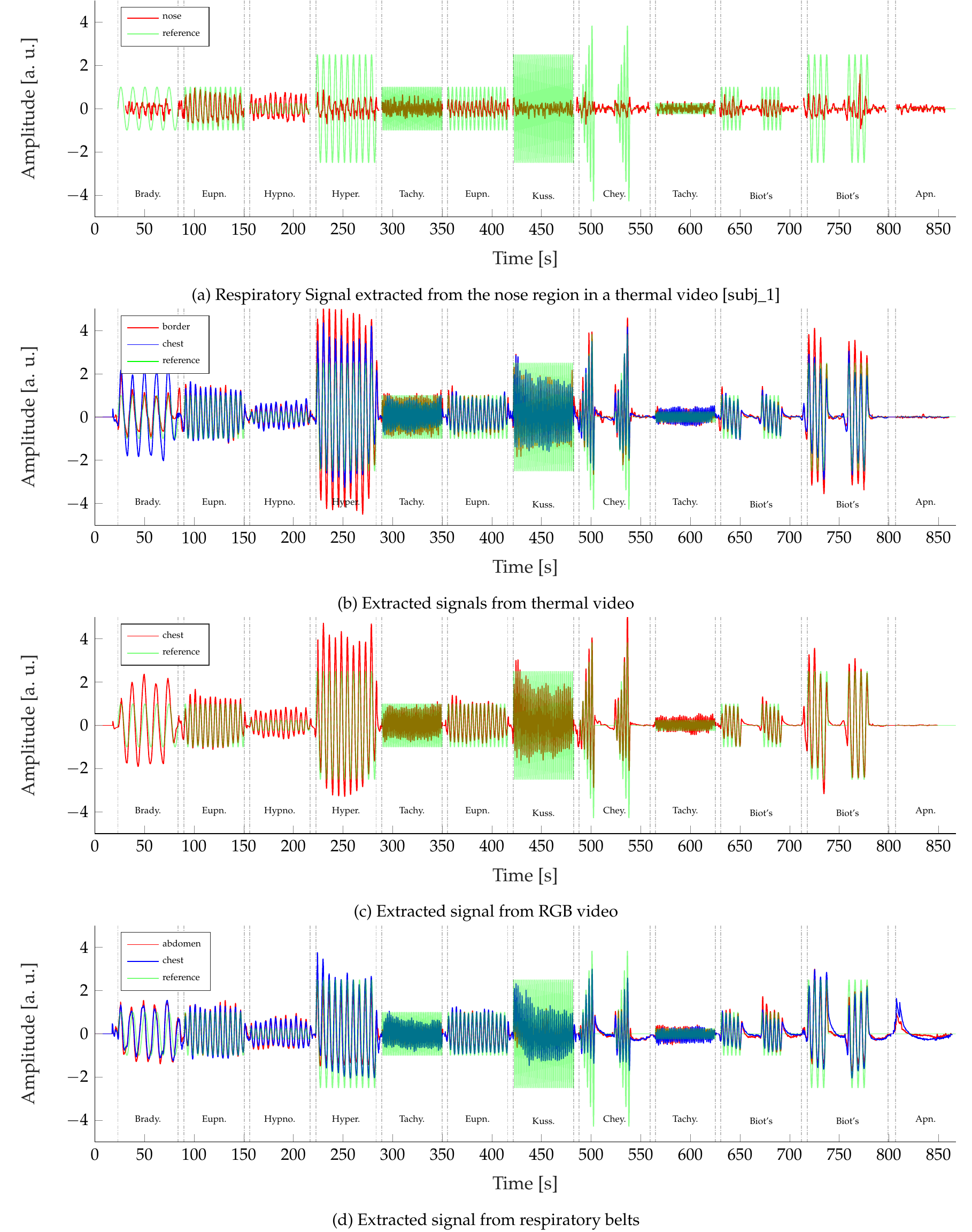}
	\caption{Temporal comparison of the mean signals of all modalities against the reference signal.} 
	\label{fig:TrackingComparison}
\end{figure} 

Figure \ref{fig:TrackingComparison} visually compares the extracted respiratory signals, grouped by their origin modality, against the reference signal, to which all subjects adjusted their breathing during the recordings. The results of the nose \gls{IRT} signal extraction approach are exemplary shown with the recordings of a single subject in Figure \ref{fig:TrackingComparison}a. It can be seen that the signal is cleary oscillating with the reference, but no significant amplitude variations are visible. For this reason, the extraction approach was not further examined as detecting amplitude differences between the breathing patterns is crucial for their correct classification. 

Figure \ref{fig:TrackingComparison}b displays the remaining respiratory signal extraction approaches for \gls{IRT} videos. The displayed signals represent the mean of all extracted signals over all recorded subjects. Regarding the frequency, an excellent agreement between both approaches and the reference signal can be observed. Clearly visible changes in amplitude during the different sections correlate well with the reference, but it can be seen that the signal amplitudes are visibly larger for the chest \gls{IRT} approach during bradypnea and for the border \gls{RoI} approach during hyperpnea. As the respiratory signals from RGB videos in \ref{fig:TrackingComparison}c were extracted in the same way as the chest \gls{IRT} signals, both are very similar to each other. Only during phases of apnea, the \gls{IRT}-based signal seems to be less noisy. The border-based \gls{IRT} signal on the other hand shows the most noise during those phases. 

Visually the best accordance with the reference show signals extracted from the respiratory chest and abdomen belts in \ref{fig:TrackingComparison}d. Here, the correlation between the signals frequency and amplitude are excellent. Only during phases of apnea, the belt-based signals performed noticeably worse, due to noticeable baseline wander. 

\subsection{Feature Extraction}
Figure \ref{fig:TemporalFeaturesView} in section \ref{sec:ArtifactCorrection} shows a high temporal resolution of the extracted respiratory features. It can be noted that the fusion of the two redundantly extracted feature sets reduces the  artifacts in the merged feature set. The accordance between the \gls{RR}s (Figure \ref{fig:TemporalFeaturesView}b) and \gls{RE}s (Figure \ref{fig:TemporalFeaturesView}c) is high, which implies a high certainty of the parameters. In signals with low \gls{RE}s, it could be noted that the \gls{CWT}-ridge based approach performed much better than the peak detection approach, as single missed peaks altered the feature extraction result significantly.  

Table \ref{tab:RMSE} shows the \gls{RMSE} for the different respiratory breathing patterns compared to the reference signal, which all subjects were requested to follow breathing during the recordings. Therefore, the \gls{RMSE} values do not only originate from the performance of the extraction algorithms alone, but are superimposed with the systematic errors the subjects made, while breathing according to the reference signal. However, since these errors influence all extraction approaches equally, the comparison between them still unveils their relative performance against each other. Regarding \gls{RMSE}, the chest belt shows the best performance. Nevertheless, the previously observed inferior performance during respiratory arrest is also reflected in large error values in Biot's respiration and apnea. The contactless modalities perform comparibly well and show only slightly higher errors than the chest belt. However, due to the extremely high errors during phases of apnea for the chest belt signals, the mean \gls{RMSE} of the contactless extracted signals outperforms the gold standard. The abdominally placed respiratory belt has the worst performance.

\begin{table}[H] 
	\centering
	\caption{Root-mean-square errors (RMSE) for the respiratory rate (\gls{RR}) extraction of the different recorded breathing patterns and modalities. The unit of all values is in \si{Breaths\per\minute}.}
	\label{tab:RMSE}
	\begin{tabular}{lccccc}
		\toprule
		\textbf{Pattern} & \multicolumn{1}{p{4.5em}}{\textbf{Chest IRT}} & \multicolumn{1}{p{4.5em}}{\textbf{Border IRT}} & \textbf{RGB} & \textbf{Abd. Belt} & \textbf{Chest Belt} \\
		\midrule
		Bradypnea & \cellcolor[rgb]{ 1,  .922,  .518}0,33 & \cellcolor[rgb]{ 1,  .914,  .518}0,44 & \cellcolor[rgb]{ 1,  .918,  .518}0,34 & \cellcolor[rgb]{ 1,  .914,  .518}0,42 & \cellcolor[rgb]{ .643,  .816,  .494}0,12 \\
		Eupnea  & \cellcolor[rgb]{ .831,  .871,  .506}0,21 & \cellcolor[rgb]{ .827,  .871,  .506}0,21 & \cellcolor[rgb]{ .816,  .867,  .506}0,20 & \cellcolor[rgb]{ .8,  .863,  .506}0,19 & \cellcolor[rgb]{ .49,  .773,  .486}0,05 \\
		Hypopnea & \cellcolor[rgb]{ 1,  .922,  .518}0,28 & \cellcolor[rgb]{ 1,  .918,  .518}0,36 & \cellcolor[rgb]{ .98,  .914,  .514}0,28 & \cellcolor[rgb]{ .902,  .89,  .51}0,24 & \cellcolor[rgb]{ .6,  .804,  .494}0,10 \\
		Hyperpnea & \cellcolor[rgb]{ .839,  .875,  .506}0,21 & \cellcolor[rgb]{ .69,  .831,  .498}0,14 & \cellcolor[rgb]{ .678,  .827,  .498}0,13 & \cellcolor[rgb]{ .894,  .89,  .51}0,24 & \cellcolor[rgb]{ .498,  .776,  .486}0,05 \\
		Tachypnea & \cellcolor[rgb]{ .906,  .894,  .51}0,24 & \cellcolor[rgb]{ .906,  .894,  .51}0,24 & \cellcolor[rgb]{ .902,  .89,  .51}0,24 & \cellcolor[rgb]{ 1,  .882,  .51}0,88 & \cellcolor[rgb]{ .612,  .808,  .494}0,11 \\
		Kussmaul & \cellcolor[rgb]{ 1,  .914,  .518}0,40 & \cellcolor[rgb]{ .98,  .914,  .514}0,28 & \cellcolor[rgb]{ 1,  .922,  .518}0,29 & \cellcolor[rgb]{ 1,  .918,  .518}0,37 & \cellcolor[rgb]{ .69,  .831,  .498}0,14 \\
		Cheyne Stokes & \cellcolor[rgb]{ 1,  .902,  .518}0,56 & \cellcolor[rgb]{ .996,  .827,  .502}1,61 & \cellcolor[rgb]{ 1,  .882,  .51}0,86 & \cellcolor[rgb]{ 1,  .882,  .51}0,88 & \cellcolor[rgb]{ .871,  .882,  .51}0,23 \\
		Biot's  & \cellcolor[rgb]{ .996,  .839,  .502}1,48 & \cellcolor[rgb]{ 1,  .906,  .518}0,52 & \cellcolor[rgb]{ 1,  .875,  .51}0,97 & \cellcolor[rgb]{ .988,  .675,  .471}3,76 & \cellcolor[rgb]{ .996,  .831,  .502}1,59 \\
		Apnea   & \cellcolor[rgb]{ .388,  .745,  .482}0,00 & \cellcolor[rgb]{ 1,  .902,  .514}0,60 & \cellcolor[rgb]{ 1,  .859,  .506}1,20 & \cellcolor[rgb]{ .973,  .412,  .42}7,43 & \cellcolor[rgb]{ .98,  .502,  .439}6,20 \\
		\midrule
		\textbf{Mean} & \cellcolor[rgb]{ .388,  .745,  .482}\textbf{0,41} & \cellcolor[rgb]{ .906,  .894,  .51}\textbf{0,49} & \cellcolor[rgb]{ 1,  .922,  .518}\textbf{0,50} & \cellcolor[rgb]{ .973,  .412,  .42}\textbf{1,60} & \cellcolor[rgb]{ .992,  .714,  .478}\textbf{0,95} \\
		\textbf{Median} & \cellcolor[rgb]{ .965,  .91,  .514}\textbf{0,31} & \cellcolor[rgb]{ .996,  .8,  .494}\textbf{0,40} & \cellcolor[rgb]{ 1,  .922,  .518}\textbf{0,32} & \cellcolor[rgb]{ .973,  .412,  .42}\textbf{0,65} & \cellcolor[rgb]{ .388,  .745,  .482}\textbf{0,13} \\
		\bottomrule
	\end{tabular}%
\end{table}

\begin{figure}[htpb]
	\centering
	\includegraphics[]{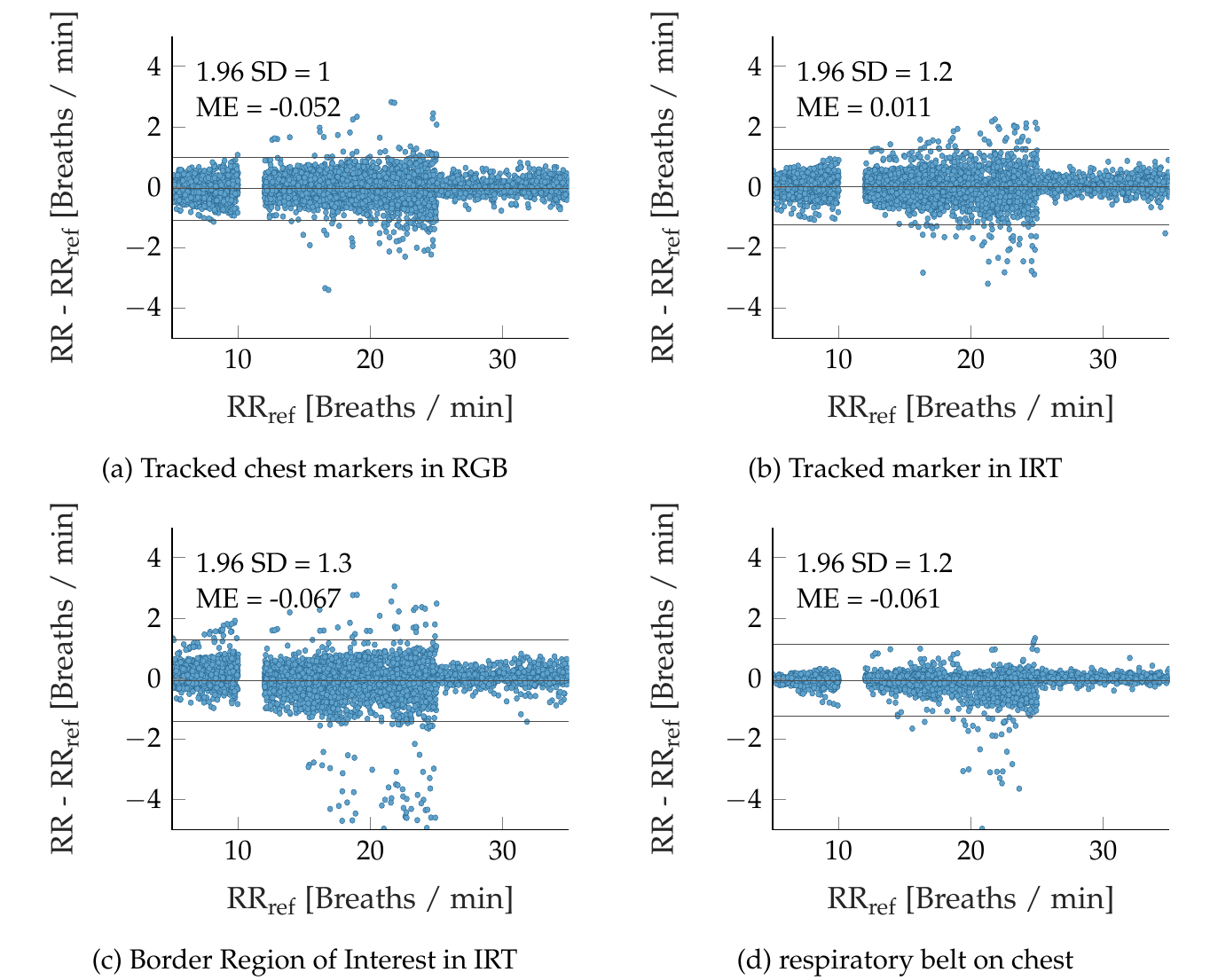}
	\caption{Bland-Altman diagrams for extracted respiratory rates.} 
	\label{fig:BAPlots}
\end{figure}

Figure \ref{fig:BAPlots} compares the contactless and the chest belt \gls{RR} extraction against the reference signal using Bland-Altman plots. The respiratory rates of the animated reference signal $RR_{\text{ref}}$ were used as ground truth. The errors during apnea phases were excluded. As in Table \ref{tab:RMSE}, it can be clearly seen that the median errors of the chest belt are significantly smaller. However, the Biot's breathing patterns between 12 and 25 \si{Breaths\per\minute} introduce only few (see Table \ref{tab:Outliers}), but big outliers, which increase the overall standard deviation. Thereby, the standard deviations of all modalities and algorithms are very similar. The marker-based approaches show that the errors from RGB-videos are more equally distributed, while the \gls{IRT}-based \gls{RR} tend to have a smaller median error, but more outliers. The border \gls{RoI} feature extraction approach shows the largest \gls{RR} mean errors of all contactless modalities and has also the highest number of outliers (see Table \ref{tab:Outliers}). Interestingly, the this approach produces the smallest median \gls{RMSE} of all modalities and performs equally well as the gold standard (the chest belt signals), when comparing the amplitude errors as can be seen in Table \ref{tab:AmplitudeErrors}. Except for patterns with high \gls{RE}s, as was seen before in Figure \ref{fig:TrackingComparison}b, the errors are lower than for all other algorithms and modalities. Chest marker tracking in thermal videos also yields low amplitude errors, except for Biot's breathing and Bradypnea. Regarding amplitude errors, the chest RGB approach had the weakest performance. 

\glsunset{RR}
\begin{table}[H] 
	\centering
	\caption{Outliers with errors greater than two standard deviations from the mean value for the respiratory rate (\gls{RR}) extraction of the different recorded breathing patterns and modalities.}
	\label{tab:Outliers}
	\begin{tabular}{lrrrrr}
		\toprule
		\textbf{Pattern} & \multicolumn{1}{l}{\textbf{Chest IRT}} & \multicolumn{1}{l}{\textbf{Border IRT}} & \multicolumn{1}{l}{\textbf{RGB}} & \multicolumn{1}{l}{\textbf{Abd. Belt}} & \multicolumn{1}{l}{\textbf{Chest Belt}} \\
		\midrule
		Bradypnea & \cellcolor[rgb]{ 1,  .914,  .518}5 & \cellcolor[rgb]{ 1,  .906,  .518}9 & \cellcolor[rgb]{ 1,  .91,  .518}7 & \cellcolor[rgb]{ 1,  .906,  .518}9 & \cellcolor[rgb]{ .388,  .745,  .482}0 \\
		Eupnea  & \cellcolor[rgb]{ 1,  .922,  .518}1 & \cellcolor[rgb]{ .388,  .745,  .482}0 & \cellcolor[rgb]{ 1,  .922,  .518}1 & \cellcolor[rgb]{ 1,  .922,  .518}2 & \cellcolor[rgb]{ .388,  .745,  .482}0 \\
		Hypopnea & \cellcolor[rgb]{ 1,  .922,  .518}1 & \cellcolor[rgb]{ 1,  .918,  .518}3 & \cellcolor[rgb]{ 1,  .918,  .518}3 & \cellcolor[rgb]{ .388,  .745,  .482}0 & \cellcolor[rgb]{ .388,  .745,  .482}0 \\
		Hyperpnea & \cellcolor[rgb]{ .388,  .745,  .482}0 & \cellcolor[rgb]{ .388,  .745,  .482}0 & \cellcolor[rgb]{ .388,  .745,  .482}0 & \cellcolor[rgb]{ 1,  .922,  .518}1 & \cellcolor[rgb]{ .388,  .745,  .482}0 \\
		Tachypnea & \cellcolor[rgb]{ .388,  .745,  .482}0 & \cellcolor[rgb]{ .388,  .745,  .482}0 & \cellcolor[rgb]{ .388,  .745,  .482}0 & \cellcolor[rgb]{ 1,  .922,  .518}1 & \cellcolor[rgb]{ .388,  .745,  .482}0 \\
		Kussmaul & \cellcolor[rgb]{ .388,  .745,  .482}0 & \cellcolor[rgb]{ .388,  .745,  .482}0 & \cellcolor[rgb]{ .388,  .745,  .482}0 & \cellcolor[rgb]{ .388,  .745,  .482}0 & \cellcolor[rgb]{ .388,  .745,  .482}0 \\
		Cheyne Stokes & \cellcolor[rgb]{ 1,  .922,  .518}2 & \cellcolor[rgb]{ 1,  .902,  .514}10 & \cellcolor[rgb]{ 1,  .922,  .518}2 & \cellcolor[rgb]{ 1,  .918,  .518}3 & \cellcolor[rgb]{ .388,  .745,  .482}0 \\
		Biot's  & \cellcolor[rgb]{ 1,  .91,  .518}7 & \cellcolor[rgb]{ 1,  .922,  .518}2 & \cellcolor[rgb]{ 1,  .91,  .518}7 & \cellcolor[rgb]{ 1,  .855,  .506}29 & \cellcolor[rgb]{ 1,  .906,  .518}9 \\
		Apnea   & \cellcolor[rgb]{ .388,  .745,  .482}0 & \cellcolor[rgb]{ 1,  .906,  .518}9 & \cellcolor[rgb]{ 1,  .894,  .514}13 & \cellcolor[rgb]{ .973,  .412,  .42}210 & \cellcolor[rgb]{ .976,  .416,  .424}209 \\
		\midrule
		\textbf{Sum} & \cellcolor[rgb]{ .388,  .745,  .482}\textbf{16} & \cellcolor[rgb]{ 1,  .922,  .518}\textbf{33} & \cellcolor[rgb]{ 1,  .922,  .518}\textbf{33} & \cellcolor[rgb]{ .973,  .412,  .42}\textbf{255} & \cellcolor[rgb]{ .98,  .498,  .439}\textbf{218} \\
		\textbf{Ratio} & \cellcolor[rgb]{ .388,  .745,  .482}\textbf{0,18\%} & \cellcolor[rgb]{ 1,  .922,  .518}\textbf{0,37\%} & \cellcolor[rgb]{ 1,  .922,  .518}\textbf{0,37\%} & \cellcolor[rgb]{ .973,  .412,  .42}\textbf{2,83\%} & \cellcolor[rgb]{ .98,  .498,  .439}\textbf{2,42\%} \\
		\bottomrule
	\end{tabular}%
\end{table}

\glsunset{RMSE}
\begin{table}[htpb] 
	\centering
	\caption{Root-mean-square errors (\gls{RMSE}) for the respiratory amplitude extraction of the different recorded breathing patterns across all modalities. The unit of all values is in normalized units.}
	\label{tab:AmplitudeErrors}
	\begin{tabular}{lccccc}
		\toprule
		\textbf{Pattern} & \multicolumn{1}{p{4.5em}}{\textbf{Chest IRT}} & \multicolumn{1}{p{4.5em}}{\textbf{Border IRT}} & \multicolumn{1}{p{4.5em}}{\textbf{RGB}} & \textbf{Abd. Belt} & \textbf{Chest Belt} \\
		\midrule
		Bradypnea & \cellcolor[rgb]{ .996,  .831,  .502}0,81 & \cellcolor[rgb]{ .761,  .851,  .502}0,26 & \cellcolor[rgb]{ .996,  .835,  .502}0,79 & \cellcolor[rgb]{ 1,  .922,  .518}0,42 & \cellcolor[rgb]{ .886,  .886,  .51}0,34 \\
		Eupnea  & \cellcolor[rgb]{ .467,  .769,  .486}0,06 & \cellcolor[rgb]{ .502,  .776,  .486}0,08 & \cellcolor[rgb]{ .514,  .78,  .486}0,09 & \cellcolor[rgb]{ .482,  .773,  .486}0,07 & \cellcolor[rgb]{ .475,  .769,  .486}0,06 \\
		Hypopnea & \cellcolor[rgb]{ .812,  .867,  .506}0,29 & \cellcolor[rgb]{ .776,  .855,  .502}0,27 & \cellcolor[rgb]{ .918,  .894,  .51}0,36 & \cellcolor[rgb]{ .933,  .902,  .514}0,37 & \cellcolor[rgb]{ .902,  .89,  .51}0,35 \\
		Hyperpnea & \cellcolor[rgb]{ .996,  .851,  .506}0,73 & \cellcolor[rgb]{ .973,  .412,  .42}2,57 & \cellcolor[rgb]{ .984,  .584,  .455}1,86 & \cellcolor[rgb]{ .996,  .8,  .494}0,94 & \cellcolor[rgb]{ .996,  .839,  .502}0,77 \\
		Tachypnea & \cellcolor[rgb]{ .89,  .886,  .51}0,34 & \cellcolor[rgb]{ .808,  .867,  .506}0,29 & \cellcolor[rgb]{ .867,  .882,  .51}0,33 & \cellcolor[rgb]{ .98,  .914,  .514}0,41 & \cellcolor[rgb]{ .812,  .867,  .506}0,29 \\
		Kussmaul & \cellcolor[rgb]{ 1,  .871,  .51}0,64 & \cellcolor[rgb]{ .98,  .486,  .435}2,26 & \cellcolor[rgb]{ .984,  .627,  .463}1,66 & \cellcolor[rgb]{ .992,  .773,  .49}1,06 & \cellcolor[rgb]{ .996,  .827,  .502}0,83 \\
		Cheyne Stokes & \cellcolor[rgb]{ .996,  .824,  .502}0,84 & \cellcolor[rgb]{ .996,  .824,  .502}0,85 & \cellcolor[rgb]{ .996,  .796,  .494}0,95 & \cellcolor[rgb]{ .992,  .753,  .486}1,14 & \cellcolor[rgb]{ .996,  .812,  .498}0,90 \\
		Biot's  & \cellcolor[rgb]{ .996,  .827,  .502}0,83 & \cellcolor[rgb]{ 1,  .894,  .514}0,54 & \cellcolor[rgb]{ .996,  .847,  .506}0,74 & \cellcolor[rgb]{ .996,  .812,  .498}0,89 & \cellcolor[rgb]{ .996,  .827,  .502}0,82 \\
		Apnea   & \cellcolor[rgb]{ .388,  .745,  .482}0,00 & \cellcolor[rgb]{ .404,  .749,  .482}0,01 & \cellcolor[rgb]{ .388,  .745,  .482}0,00 & \cellcolor[rgb]{ .412,  .749,  .482}0,02 & \cellcolor[rgb]{ .4,  .749,  .482}0,01 \\
		\midrule
		\textbf{Mean} & \cellcolor[rgb]{ .494,  .773,  .486}\textbf{0,50} & \cellcolor[rgb]{ .973,  .412,  .42}\textbf{0,79} & \cellcolor[rgb]{ .98,  .51,  .439}\textbf{0,75} & \cellcolor[rgb]{ 1,  .922,  .518}\textbf{0,59} & \cellcolor[rgb]{ .388,  .745,  .482}\textbf{0,49} \\
		\textbf{Median} & \cellcolor[rgb]{ .996,  .784,  .494}\textbf{0,57} & \cellcolor[rgb]{ .388,  .745,  .482}\textbf{0,42} & \cellcolor[rgb]{ .973,  .412,  .42}\textbf{0,75} & \cellcolor[rgb]{ 1,  .922,  .518}\textbf{0,51} & \cellcolor[rgb]{ .408,  .749,  .482}\textbf{0,42} \\
		\bottomrule
	\end{tabular}%
\end{table}

\subsection{Dataset Analysis}
\begin{figure}[H]
	\centering
	\includegraphics[]{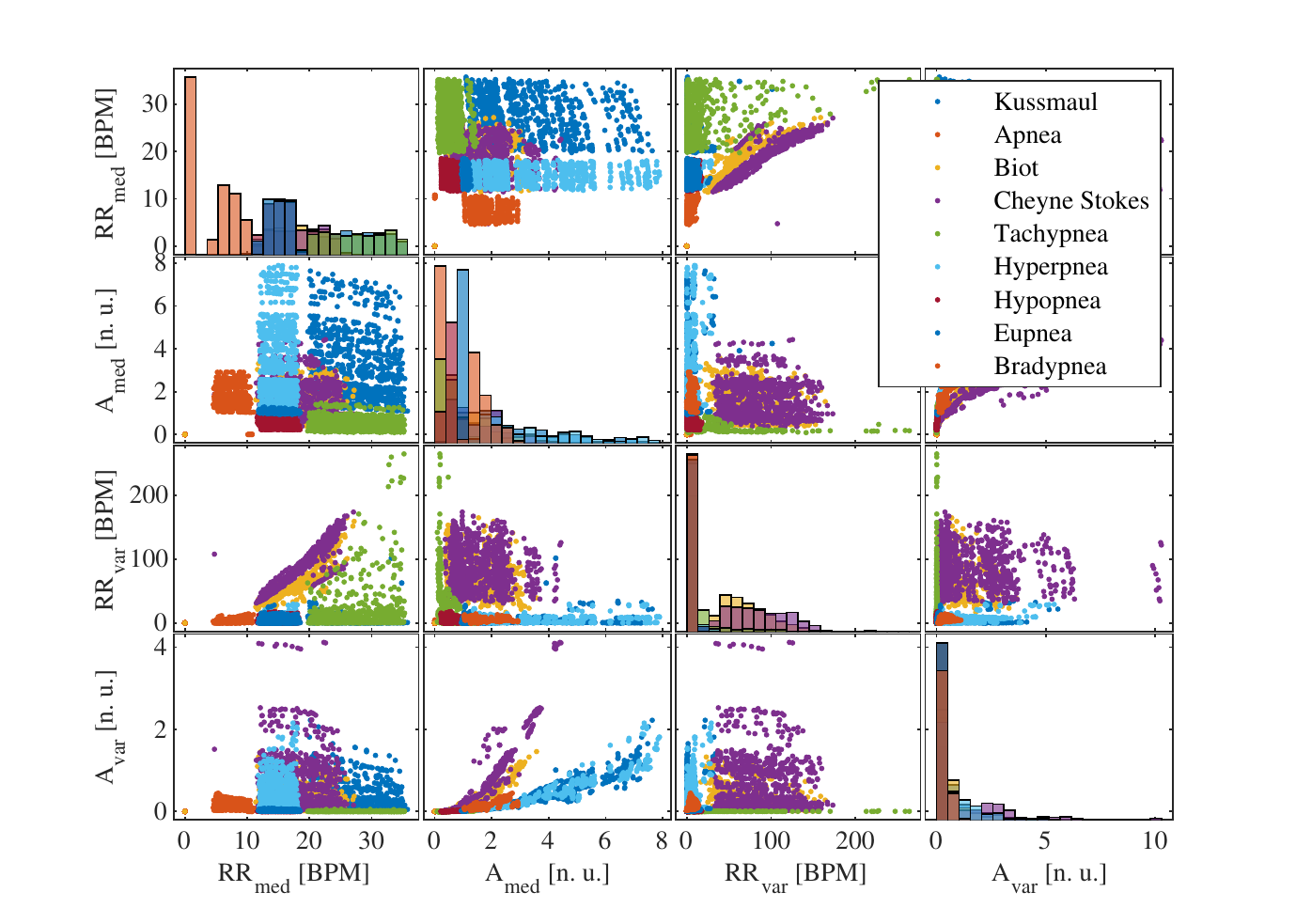}
	\caption{Scatter plots for all extracted respiratory feature combinations.} 
	\label{fig:Plotmatrix}
\end{figure}

Figure \ref{fig:Plotmatrix} shows a plot matrix with the scatter plots of all extracted features by means of their median value over a complete breathing pattern segment. The respiratory width was excluded, as it did not provide a significant contribution to the separation of the different respiratory pattern. Generally, no distinct lines can be observed in the feature distributions of $RR_{\text{med}}$ and the signal amplitudes $A_{\text{med}}$. Instead, they are evenly distributed over the entire defined range of values. This confirms the effectiveness of the chosen data augmentation principles and prevents overfitting of classifiers at systematic gaps that would otherwise have arisen in the feature value ranges due to the study design. 

The histograms on the main diagonal of the plot matrix show the distribution of all classes along their complete value range. The \gls{RR} is clearly the feature, which alone can provide the best separation between the displayed pattern classes. However, it is only sufficient for classification between apnea, bradypnea, eupnea and tachypnea. All other classes need multiple features for their extraction. Here, the most important pair of features appear to be the median \gls{RR} $RR_{\text{med}}$ and signal amplitude $A_{\text{med}}$, which show the \gls{RE}. By adding this feature, also hypopnea, hyperpnea and Kussmaul breathing become seperable. For complex respiratory patterns, like Biot's and Cheyne-Stokes breathing, which cover a wide range of \gls{RR}s and \gls{RE}s, some separation between the other patterns can be established by introducing the variance features. Still, optically no clear seperation can be achieved for these patterns.

\subsection{Classifcation Results}
\glsunset{RMSE}
The total classification for the different accuracies extraction algorithms and modalities are shown in Table~\ref{tab:Accuracies}. With almost \SI{96}{\percent} accuracy, the chest marker tracking approach in \gls{IRT}-videos outperformed the other approaches and even the chest belt, which serves as gold standard. Interestingly, all contactless approaches outperformed the gold standard, although it showed equal or better performance during the feature extraction. 

\begin{table}[htpb] 
	\centering
	\caption{Breathing pattern classification results for all extracted signals using a one vs. one multiclass Support Vector Machine}
	\label{tab:Accuracies}
	\begin{tabular}{lrrrrr}
		\toprule
		& \multicolumn{1}{p{4.145em}}{\textbf{Chest IRT}} & \multicolumn{1}{p{4.645em}}{\textbf{Border RoI}} & \multicolumn{1}{p{3.645em}}{\textbf{Chest RGB}} & \multicolumn{1}{c}{\textbf{Abd. Belt}} & \multicolumn{1}{c}{\textbf{Chest Belt}} \\
		\midrule
		\textbf{Accuracy} & 95,79\% & 94,93\% & 93,48\% & 91,87\% & 92,37\% \\
		\bottomrule
	\end{tabular}%
\end{table}

\begin{figure}[htpb]
	\centering
	\includegraphics[]{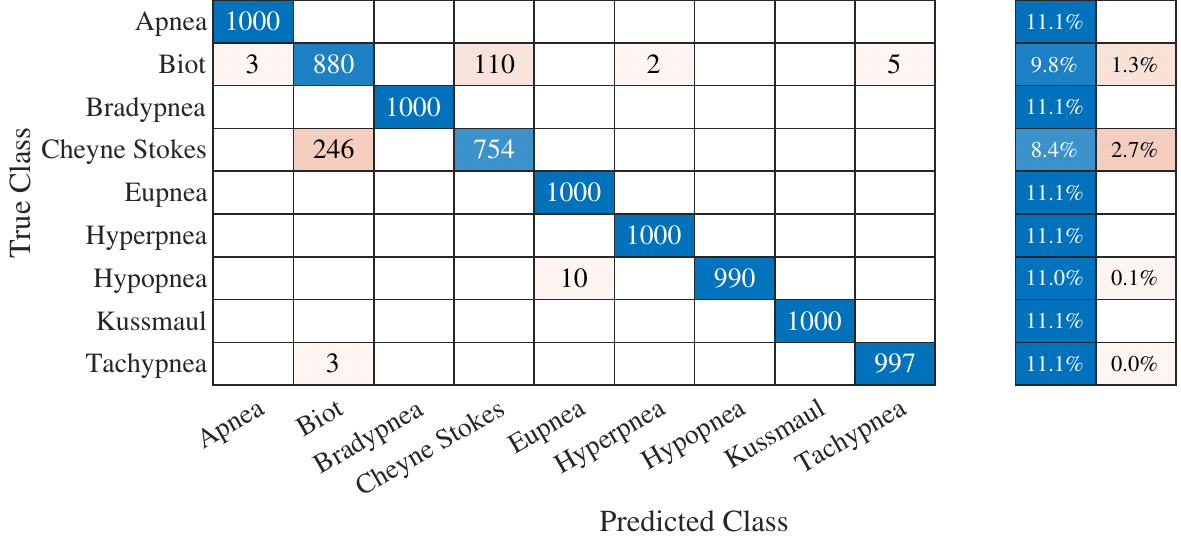}
	\caption{Confusion matrix for the one vs. one multiclass Support Vector Machine breathing pattern classification of the respiratory signal extracted by chest feature tracking in thermal videos} 
	\label{fig:ConfutionMatrix}
\end{figure}

Figure~\ref{fig:ConfutionMatrix} shows the confusion matrix of the respiratory signal extracted by chest feature tracking in \gls{IRT}-videos. It can be seen that the results are almost perfect. The biggest source of confusion originates from the complex respiratory patterns (Biot's breathing and Cheyne Stokes respiration). With confusion rates of \SI{1.3}{\percent} and \SI{2.7}{\percent}, these breathing patterns were more than thirteen times more likely to be misclassified than hypopnea, the following largest source of confusion with \SI{0.1}{\percent}.

\section{Discussion}
This work presented several approaches to extract and classify respiratory signals from RGB, or \gls{IRT} videos. Their performances were compared to respiratory belt sensors, located at chest and abdomen, which served as gold standards. With the chest \gls{IRT} approach, \SI{95.79}{\percent} classification accuracy was achieved. The other contactless approaches, border \gls{RoI} and chest RGB performed comparably well and all outperformed the gold standard. These results are promising and clearly show the feasibilty for contactless systems to  perform a complete respiratory pattern assessment. As mentioned in the introduction, the knowledge over the current respiratory pattern can not only give insight into the patients gas exchange, but can also be used as indicator for a broad range of medical conditions. Therefore, the relevance of a system, which is capable of continuously classifying a patient's respiratory pattern in an unobstrusive and cost-effecive manner should not be underestimated. 

Still, it should still be discussed which of the presented approaches is the most promising for a timely product development, where the performance differences in the different approaches come from, and which improvements, respectively extensions in the field of automated respiration analysis would be conceivable in the future. During evaluation, interesting differences regarding the quality of respiratory amplitude or \gls{RE} extraction became apparent. Their reasons shall be discussed in the following: In contrast to the other modalities and measurement locations, the nose \gls{IRT} signal was not able to show any differences in \gls{RE} over the different breathing patterns. This is due to the fact that the temperature of the in- and exhaled air, which leads to temperature changes in the \gls{IRT} recordings, does not vary with the \gls{RE}, but with the respiratory flow. This renders it useless for the presented classification approach, but makes it very interesting for more advances respiratoy assessment approaches. For example, the monitoring of the relationship between flow and effort yields information about a lung's mechanical condition in terms of resistance and compliance. These two measures are vitally important when diagnosing obstructive, or restrictive lung diseases. The signals from the border RoI approach were inferior, compared to the chest-tracker-based approaches. This is due to the fact that for the latter approach more of the extracted raw signals contain relevant information, as they are all located on the subjects chest. Only around the shoulder regions, the border \gls{RoI}s deliver signals with relevant respiratory content. This also explains the observed amplitude differences between the two approaches: while the chest-tracker-based approaches mainly evaluate chest rise, the border \gls{RoI} approach focuses on shoulder movement. Healthy subjects increase their lung volume during phases of apnea (e.g. during meditation) to maintain a constant minute ventilation as compared to eupnea. This results in a larger chest excursion and thus larger signal amplitudes for the chest-tracker-based approaches. On the other hand, the border \gls{RoI} approach is highly sensitive to the utilization of auxiliary respiratory muscles, which result in an additional lift of the shoulders. Hence, the border \gls{RoI} signals had a significantly higher amplitude during phases with high \gls{RE}. By simultaneously using the chest signals as surrogate for lung volume and the shoulder signals as surrogate for high \gls{RE}s, an even more fine-grained automatic assessment could be provided. 

Also regarding the measurement modalities, there were notable differences, which shall be addressed in the following: In general, \gls{IRT} has proven to be superior, compared to the other modalities. This can be explained by a high contrast of the subject against the colder background, but also on the subjects chest due to wrinkles in the clothing (see \ref{fig:RoIs}). Although RGB recordings have a higher resolution, the low contrast on single-colored textiles reduces the quality of the feature point tracking algorithm. Furthermore, over- and underexposure can further limit the usable contrast. However, there may exist scenarios, where \gls{IRT} has less contrast than RGB sensors (e.g. outside during hot or cold weather, when either the background has a similar temperature as the subject, or the subject wears insulated clothing). Because of the low cost of RGB sensors it is therefore proposed to use both modalities simultaneously and to perform signal fusion to further enhance the robustness. Chest belt sensors have shown to deliver high quality signals. However, during apnea, they were prone to artifacts due to baseline wander. In contrast to all other modalities, only one signal source was available and no implicit detrending through the \gls{PCA} fusion algorithm could be performed. This fact contributed to the inferior performance during phases of apnea. 

For breathing pattern classification, the contactless approaches outperformed the respiratory belts, although the chest belt signal showed better performance for \gls{RR} and \gls{RE} extraction. This can not alone be explained by the fact that respiratory belt signals showed inferior performance during phases of apnea. This fact contributed \SI{0.1}{\percent} of false classifications to the overall performance. The main difference can be found in the error rates for classification of complex breathing patterns (Biots’s breathing and Cheyne-Stokes respiration). Here, the contactless approaches performed better. A possible explanation might be that the contactless approaches tended to exaggerate high \gls{RE}s, which might me beneficial for classification. Nevertheless, because complex respiratory patterns consist of multiple simple patterns, one-shot classifiers, like the one presented in this manuscript are not perfectly suited for their classification. By using more advanced classifiers, there is the possibility that all discussed modalities yield the same results. 

Multiple approaches could be tested to also accurately classify complex breathing patterns. The first possibility that the authors can think of would to use the \gls{CWT} spectrograms as inputs for a suitable 2D Convolutional Neural Network. This would provide the classifier with all relevant time-frequency and amplitude information, necessary for successful classification. Also, the temporal feature signals themselves could be used as input for a suitable recurrent neural network. The memory effect of these neural networks make them ideal for classifying temporal patterns. A last possibiltiy would be to use the presented classifier in this paper only for the simple brathing patterns and to rely on a simple, downstream recurrent neural network to detect and classify more complex breathing patterns based on its outputs. 

That other approaches for complex breathing pattern classification can be contributed by other groups and that they can be compared in a standardized way, the presented data set is published. Expanding the data set to include other breathing patterns that occur, for example, in obstructive, or restrictive lung diseases, would also be an interesting prospect for subsequent work, which could further increase the medical usability of the classification results. In a larger patient study in the recovery room of the RWTH Aachen University Hospital, the results of the presented subject study are currently being compared and validated with actual pathological breathing patterns. This shall ensure that both the data set and the developed classifiers are suitable for practical use.

\vspace{6pt} 



\textbf{Funding: }This research was funded by the German Federal Ministry of Education and Research grant number 13N14772.

\textbf{Ethics: }The experiments were conducted with the help of volunteering colleagues. According to the ethics committee of the medical faculty of the RWTH Aachen University, Germany, therefore, no ethics approval was necessary. Informed consent was obtained from all subjects involved in the study.

\textbf{Acknowledgments: }The authors thank all colleagues, which volunteered in this study for their time and their excellent breathing patterns.

\textbf{Conflicts of Interest: }The authors declare no conflict of interest.
	
\textbf{Abbreviations: }The following abbreviations are used in this manuscript:
\begin{adjustwidth}{1cm}{}
	\begin{multicols}{2}[]
		\printglossary[type=\acronymtype]
	\end{multicols}
\end{adjustwidth}

\vspace{2cm}

\appendix
\section[\appendixname~\thesection]{Experimental setup}

\begin{tabularx}{\textwidth}{cccl}
	\caption{Devices used for data caption} \label{tab: caption_devices} \\
	
	\hline \textbf{Device}	& \textbf{Product Name} & \textbf{Manufacturer}	& \textbf{Information}\\ 	\hline
	
	Chest Sensor 		& PVDF Effort Sensor		& SleepSense 		& \\
	Adbodem Sensor      & PVDF Effort Sensor 		& SleepSense 		& \\
	A/D Converter		& NI USB-6008				& Nordic Instruments& 	{\def\arraystretch{1}\begin{tabular}[c]{@{}l@{}}resolution: 12 bit,\\ sampling rate: 5 kHz\end{tabular}} \vspace{10pt}\\
	\vspace{10pt}
	Ring Light          & SL-480                    & Dörr              & {\def\arraystretch{1}\begin{tabular}[c]{@{}l@{}}color temperature: 4600 K, \\ illumination power: 1000 lx\end{tabular}}                                                                             \\
	\vspace{10pt}
	RGB-Camera          & Mako G234C                & Allied Vision     & {\def\arraystretch{1}\begin{tabular}[c]{@{}l@{}}resolution: 1920 x 1080 px,\\ color depth: 8 bit,\\ frame rate: 15 FPS\\ lens: Kowa LM50HC\end{tabular}} \\
	Thermal camera	    & Boson 640                 & FLIR              & {\def\arraystretch{1}\begin{tabular}[c]{@{}l@{}}resolution: 640 x 512 px,\\ color depth: 16 bit,\\ frame rate: 30 FPS,\\ lens: 12\si{\degreeCelsius} FOV\\ spectral range: 7,5 - 13,5 \si{\micro\meter}\end{tabular}} \\
	\hline
\end{tabularx}

\begin{table}[H]
	\centering
	\caption{Pattern description of the breath-protocol, shown in Figure~\ref{fig:breathPatterns}}
	\label{tab:breathPatterns}
	\begin{tabular}{ccccc} 
		\toprule
		\textbf{No.}  & \textbf{Pattern}          & \begin{tabular}[c]{@{}c@{}}\textbf{Frequency}\\\si{Breaths\per\minute}\\ \end{tabular} & \begin{tabular}[c]{@{}c@{}}\textbf{Effort}\\norm. units units\\ \end{tabular}  & \begin{tabular}[c]{@{}c@{}}\textbf{Targ. Range}\\\si{Breaths\per\minute}\\ \end{tabular} \\ 
		\midrule
		1             & Bradypnea                 & 5                                                                  & 1            &  5 - 10\\
		2             & Eupnea                    & 12                                                                 & 1            &  12 - 18\\
		3             & Hypopnea                  & 12                                                                 & 0.25         &  12 - 18\\
		4             & Hyperpnea                 & 10                                                                 & 2.5          &  12 - 18\\
		5             & Tachypnea                 & 35                                                                 & 1            &  20 - 35\\
		6             & Eupnea                    & 15                                                                 & 1            &  12 - 25\\
		7             & Kussmaul breathing        & 30                                                                 & 2.5          &  20 - 35\\
		8             & Cheyne-Stokes respiration & 0, 20                                                             & 0-4.5         &  12 - 25\\
		9             & Tachypnea                 & 35                                                                 & 0.25         &  20 - 35\\
		10            & Biot's breathing          & 0, 15                                                              & 1            &  12 - 25\\
		11            & Biot's breathing          & 0, 10                                                             & 2.5           &  12 - 25\\
		12            & Apnea                     & 0                                                                  & 0            &  0\\
		\bottomrule
	\end{tabular}
\end{table}

\section[\appendixname~\thesection]{Used Parameters}
\begin{table}[H]
	\centering
	\caption{Parameters for Savitzky Golay pre-processing filter}
	\label{tab:SgolayParams}
	\begin{tabular}{cc} 
		\toprule
		\textbf{Order} & \textbf{Frame length} \\
		\midrule
		2 & \SI{1.5}{\second} \\
		\bottomrule
	\end{tabular}
\end{table}

\begin{table}[H]
	\centering
	\caption{Parameters for the Peak detection algorithm}
	\label{tab:FindPeaksParamas}
	\begin{tabular}{ccc} 
		\toprule
		\textbf{Prominence} & \textbf{Min. Distance} & \textbf{Max. Peak width}\\
		\midrule
		0.3 n.u. & $\frac{\SI{60}{\second\per\minute}}{\SI{40}{Breaths\per\minute}} = \SI{1.5}{\second}$ & $\frac{2}{3}\frac{\SI{60}{\second\per\minute}}{\SI{4}{Breaths\per\minute}} = \SI{10}{\second}$\\
		\bottomrule
	\end{tabular}
\end{table}

\begin{table}[H]
	\centering
	\caption{Parameters for the \glsfirst{CWT}}
	\label{tab:CWTParamas}
	\begin{tabular}{cccc} 
		\toprule
		\textbf{Wavelet} & \begin{tabular}[c]{@{}c@{}}\textbf{Time-Bandwidth}\\\textbf{product}\\ \end{tabular} & \begin{tabular}[c]{@{}c@{}}\textbf{Voices per}\\\textbf{octave}\\ \end{tabular} & \textbf{Frequency range}\\
		\midrule
		Morse & 10 & 48 & $[4 40]\si{Breaths\per\minute} = [0.067 0.667]\si{\hertz}$ \\
		\bottomrule
	\end{tabular}
\end{table}

\bibliographystyle{unsrtnat}
\bibliography{references}  

\end{document}